\documentclass[twocolumn,prb,showpacs,preprintnumbers,amsmath,amssymb]{revtex4}

\usepackage{graphicx}
\usepackage{dcolumn}
\usepackage{bm}

\usepackage{color} 

\begin{document}

\preprint{APS/123-QED}

\title{Local Semiconducting Transition in Armchair Carbon Nanotubes: 
The Effect of Periodic Bi-site Perturbation 
on Electronic and Transport Properties of Carbon Nanotubes}

\author{M. J. Hashemi}
\author{K. S\"{a}\"{a}skilahti}
\author{M. J. Puska}

\affiliation{%
Department of Applied Physics, Aalto University, P.O. Box 11100, 
FI-00076 AALTO, Finland
\\
}
\date{\today}

\begin{abstract}
In carbon nanotubes, the most abundant defects, caused for example by 
irradiation or chemisorption treatments, are small perturbing clusters 
,i.e. bi-site defects, extending over both A and B sites. The relative 
positions of these perturbing clusters play a crucial role in determining 
the electronic properties of carbon nanotubes.
Using bandstructure and electronic transport calculations, we find out that
in the case of armchair metallic nanotubes 
a band gap opens up when the clusters fulfill a certain periodicity condition.
This phenomenon might be used in future nanoelectronic devices in which
certain regions of single metallic nanotubes could be turned to
semiconducting ones. Although in this work we study specifically the 
effect of hydrogen adatom clusters, the phenomenon is general for different 
types of defects. Moreover, we study the influence of the length and 
randomness of the defected region on the electron transport through it.

\end{abstract}

\maketitle

\section{Introduction} 
Carbon based materials, carbon nanotubes and graphene, are considered
as the most promising candidates for many future technological applications 
because of their unique electronic, mechanical and optical 
properties.\cite{Anantram06} 
In the case of single wall carbon nanotubes (SWCNTs), the chirality and 
diameter determine if a SWCNT is 
metallic or semiconducting\cite{Dresselhaus}. In particular, armchair 
nanotubes with the $(n,n)$ -type chiral vectors are metallic, enabling 
their use as ultimate leads in nanoelectronics. 

Assuming the possibility of chirality-sensitive selection of nanotubes 
the first step towards single-SWCNT nanoelectronic devices is the 
formation of rectifying metal-semiconductor junctions. Occasionally,
this kind of junctions are realized due to pentagon-heptagon defects
making a seamless junction between nanotubes of different chiralities
\cite{Service96,Yao99}. More intentionally, nanotubes of different character 
can also be electron-beam welded at elevated temperatures to form 
X-, T-, and Y-junctions \cite{Terrones00}. 
Moreover, recently Lee {\em et al.} \cite{Lee08} suggested 
the joining of different types of nanotubes with covalent peptide linkages. 
Another route is to modify the electronic bandstructure 
of a single SWCNT spatially by functionalization with defects, adatoms or
molecules, or by controlled deformation. The use of modulating (saturation) 
hydrogen adsorption \cite{Gulseren03} or radial deformation \cite{Kilic00} 
has been suggested to be used to create quantum well structures for charge 
carriers. As a potential method, Wall and Ferreira \cite{Wall07} modeled 
also the effects of helical wrapping of polymeric molecules around SWCNT's. 

\begin{figure}
\includegraphics[scale=0.113,clip]{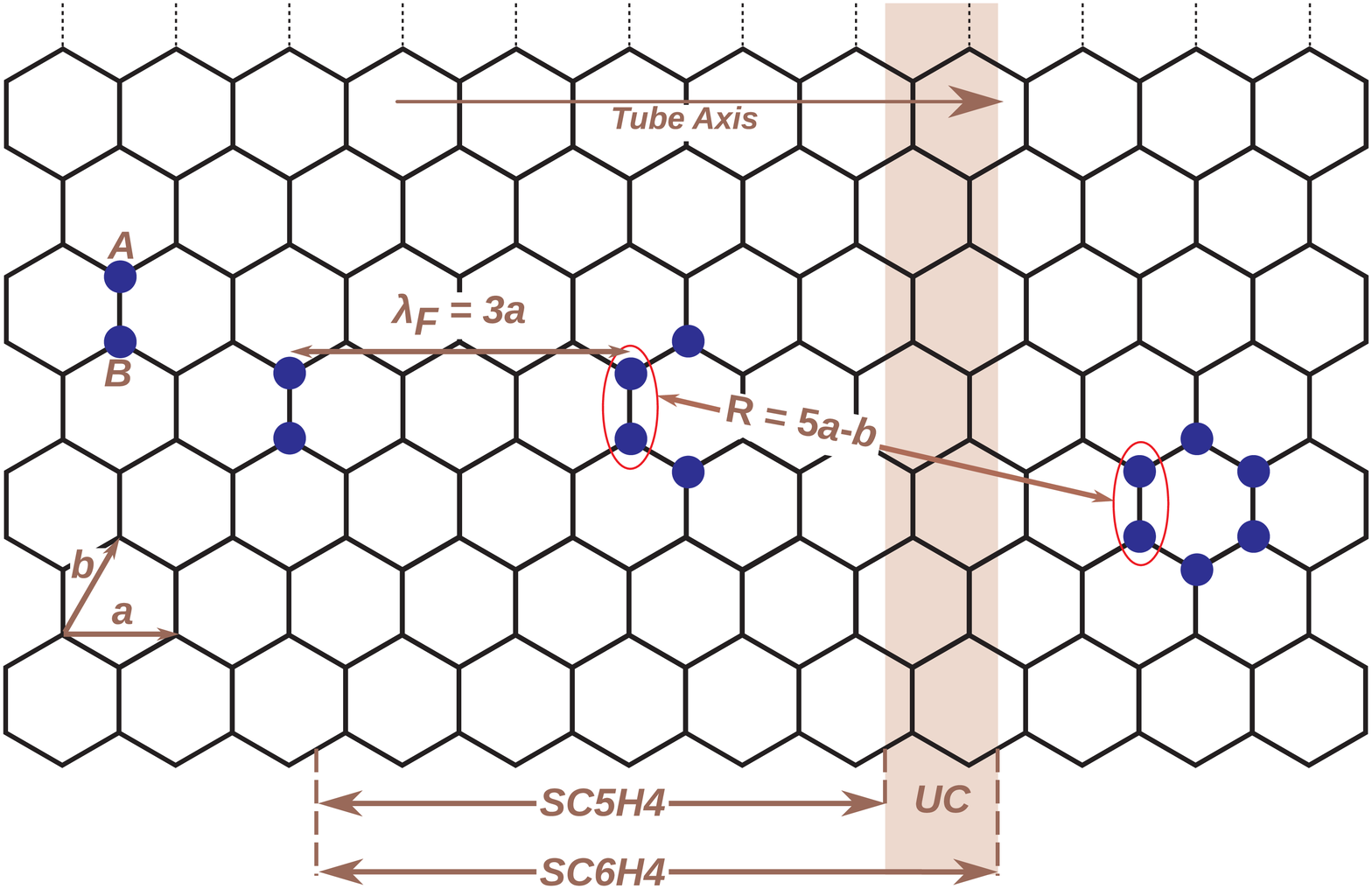}
\caption{\label{fig:nanotube} (Color online) 
Three different hydrogen clusters (blue circles) used in the calculations. 
They are shown on a piece of an armchair SWCNT's surface.
The unit vectors of the graphene sheet ($\vec{a}$ and $\vec{b}$), the A and B sublattices,
the Fermi wavelength ($\lambda_F = 3a$), and a vector connecting the adjacent 
hydrogen clusters are also given. The unit cell (UC) of the
pristine tube, as well as superlattice unit cells corresponding 
to two different periodicity of the hydrogenated SWCNTs (SC5H4 and SC6H4) are shown.
}
\end{figure}

Imperfections, their causes and effects in SWCNTs have been a subject 
of detailed studies for a long time 
\cite{PhysRevLett.84.2917, JPSJ.68.716, RevModPhys-nanotube, krasheninnikov}. 
For instance, point-like vacancies, defects and adsorbed atoms are known 
as a source of reduction in the electronic transmission through nanotubes 
\cite{PhysRevLett.92.256805, JPSJ.74.777,symmetryEffect, Wannier, partovi, InterferenceEffects}. 
The carbon atoms of a nanotube belong to two sublattices, A and B, and then 
in the case of more than one adsorbate the effect on the transmission 
strongly depends on whether the adsorbates are on the same sublattice, A or B, 
or on the different ones, A and B, as well as on their relative 
positions \cite{main, InterferenceEffects, symmetryEffect}.
In a recent study, Garc\'\i{}a-Lastra {\em et al.} \cite{main} showed that 
the relative position dependence obeys a certain rule if the adsorbates are on the same sublattice,
while they did not find any trend for adsorbates on different sublattices.
Considering the difficulty of realizing such site-selective perturbations
and the abundance of extended bi-site perturbations in experiments, it is also important to 
understand how defect clusters (collectively) affect the 
electronic properties of carbon nanotubes.

In this article we discuss the effect of perturbing defect clusters spanning 
both the A and B sublattices and we refer to them as bi-site perturbances 
or bi-site defects. In particular we discuss how undoped 
armchair SWCNTs can locally be turned from metallic to semiconducting by 
periodic bi-site perturbations. We study the effects of repeated small 
hydrogen clusters (Fig.~\ref{fig:nanotube}) on SWCNTs, but our main qualitative 
results are valid also for arbitrary bi-site perturbance. We discuss how 
the electron transport through armchair SWCNTs is affected by bi-site 
perturbations of different number, strength and periodicity along the SWCNT. Our 
calculations for the metallic armchair SWCNTs show that when
a certain rule for the relative positions of the defect clusters
is fulfilled, an energy gap around the Fermi level opens gradually by 
increasing the length of the periodically defected
region. We study the robustness of this effect against
randomness in the cluster size and geometry as well as in their relative
positions. Hydrogen clusters are realistic defect candidates because 
calculations and experiments show that adsorbed hydrogen atoms tend to cluster
on SWCNTs' sidewalls \cite{clustering, PhysRevB.75.075420, Khazaei20093306}.
The rapid development in pattern making, e.g. using block copolymer 
nanolithography \cite{patternmask, graphenenanomesh}, and in fine tuning 
and manipulating structures
even in the \AA ngstrom scale \cite{angstromSTEM} may make regular
defect systems feasible and relevant also in the experimental and practical
sense.

The organization of the present article is as follows. In Sec. II we present
the systems studied and their notation as well as the methods used in
electronic structure and transport calculations. Sec. III comprises our
results and their discussion. Sec. IV is a short summary.

\begin{figure}
\includegraphics[scale=0.62,clip]{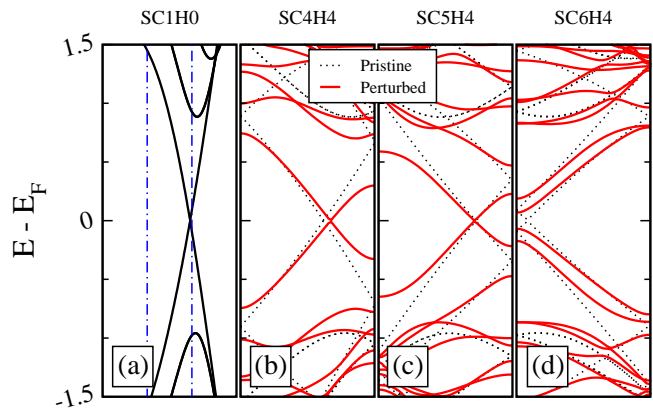}
\caption{\label{fig:bandstructure} (Color online) Effect of the 
periodically-repeated H4 clusters on the band structure of the
(8,8) SWCNT. The band structure of (a) the pristine nanotube 
SC1H0 is compared to those for nanotubes with H-atom clusters and different
supercell lengths, i.e., for (b) SC4H4, (c) SC5H4, and (d) SC6H4 (For the
notation see the text and Fig.~\ref{fig:nanotube}). The dotted lines 
in (b) - (d) denote the band structure of the pristine nanotube folded 
according to the length of the supercell. The (blue) dashed lines in panel (a) 
are band or Brillouin zone folding lines for SC3H0. For any 
SC(3*M) with an integer M, these lines are two of the 3M-1 folding lines. 
Therefore in all of these cases, the Fermi point is placed, after the folding,
near the $\Gamma$-point.}
\end{figure}

\section{Systems and Methodology} 
First we perform a set of supercell band-structure calculations in which 
supercells consist of a certain number of unitcells of a pristine 
armchair SWCNT and four hydrogen atoms adsorbed on neighboring 
carbon atoms as shown in Fig.~\ref{fig:nanotube}. In particular, we choose 
the (8,8) nanotube which has a diameter close to that often seen in 
experiments. We adopt, for example, the notation SC5H4 for the supercell 
comprising five (8,8) SWCNT unitcells (SC5) and an adsorbed cluster of 
four hydrogen atoms (H4) as depicted in Fig.~\ref{fig:nanotube}. The 
essential criterion for these clusters is that they have to perturb both 
the A and B sublattices in a plane perpendicular to the tube axis. Similar 
clusters with an odd number of hydrogen atoms would lead to similar effects.
Because we are interested in the main qualitative results and want to 
avoid the complication due to spin effects we neglect these odd-atom 
clusters in our studies.

The electronic structures are solved with the density functional theory
(DFT) within the PBE generalized gradient approximation \cite{PBE} for the
electron exchange and correlation. We employ the SIESTA package \cite{siesta} 
with non-local norm-conserving pseudopotentials \cite{TroullierMartins}
and a double-zeta plus polarization (DZP) atomic orbital basis set. 
The atomic geometry and the supercell size are relaxed until all atomic 
forces are less than $0.02$~\textit{eV/\AA }. C-C and C-H bond lengths 
in a close agreement with experiments are obtained. 

Thereafter structures for the transport calculations are made of two
semi-infinite pristine (8,8) SWCNT leads and a central region with a varying 
content. In order to see how prolongation 
of the periodically defected region changes the influence of hydrogen 
clusters on the electronic transport of nanotubes, we consider central
regions consisting of varying numbers of the different, above-mentioned 
supercells. For 
example, we may have ten SC5H4 supercells which is denoted as 
the 10(SC5H4) scattering region. Finally, we have at each end
of the central region a unitcell of the (8,8) SWCNT nanotube to ensure 
non-reflecting semi-infinite leads.

\begin{figure*}
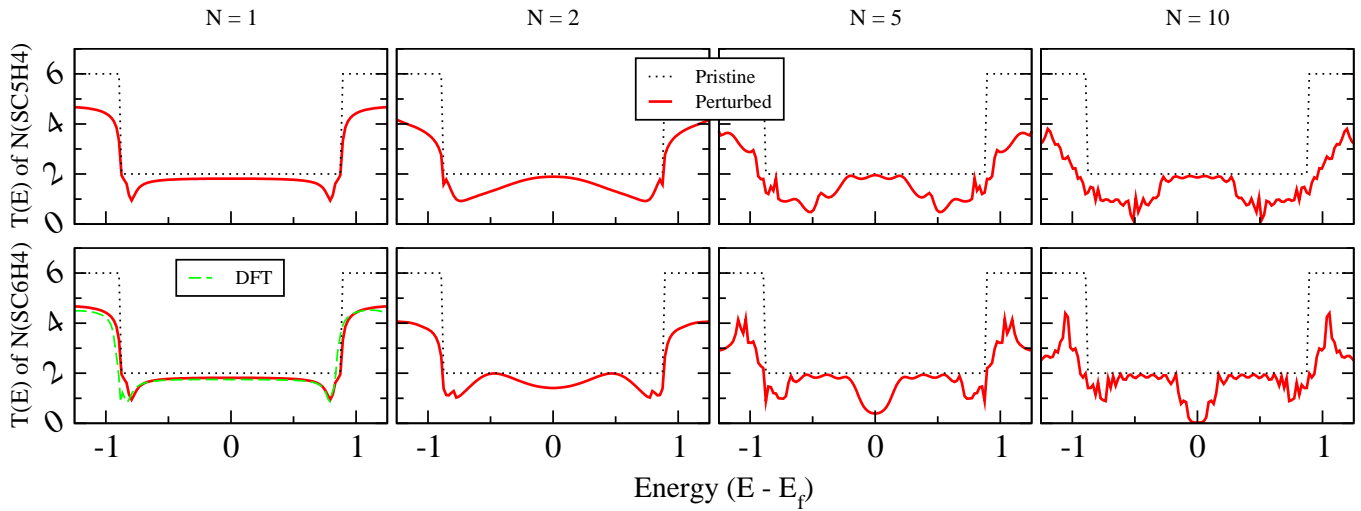

\includegraphics[scale=0.72,clip]{CNT5_4.eps}
\includegraphics[scale=0.72,clip]{CNT6_4.eps}
\caption{\label{fig:transport} (Color online) Effect of the relative
positions and the number of bi-site perturbations on the transmission
coefficient of armchair SWCNTs. The (red) solid curves in the upper and lower 
rows show the tight-binding results for central regions N(SC5H4) and N(SC6H4), 
respectively. From left to right, N=1, 2, 5 and 10.
The dotted black lines give the pristine transmission function.
The dashed (green) curve in the lower left panel gives the DFT result
calculated by the Transiesta program.}
\end{figure*}

Our transport calculations are done using the Landauer-B\"{u}ttiker
formalism. We construct the Hamiltonian matrix using the tight-binding
method with nearest-neighbor hopping. Because carbon atoms take part in
the transport process by the $p_{z}$ orbitals, the carbon atoms using
them in bonding with on-top adsorbed hydrogen atoms  will not
participate in the transport process anymore.
In our tight-binding transport calculations, we simulate this effect 
by removing the host carbon atoms binding to hydrogen atoms. For benchmarking 
we compare the transmission functions of a single SC6H4 supercell 
obtained by the tight-binding method with that calculated by the 
DFT  Transiesta program \cite{transiesta}. When the hopping parameter,
$t$, has the value of 2.3 eV we find a very good agreement
particularly around the Fermi level (See the lower left panel in 
Fig.~\ref{fig:transport}).

\section{Results and Discussion} 
Fig.~\ref{fig:bandstructure} shows
our DFT results for the band structures of the pristine (8,8) 
SWCNT as well as those of (8,8) nanotubes decorated periodically
with H4 clusters. The band structures are shown with the 
increasing supercell length, i.e., they correspond to the SC1H0, SC4H4, 
SC5H4, and SC6H4 supercells. The increase of the supercell length folds 
the band structure of the pristine single-unit-cell nanotube as shown by 
dotted lines in panels (b) - (d). There are qualitatively two different 
repeating bandstructure schemes with respect to the band crossing point 
at the Fermi energy represented by the SC4H4 or SC5H4 supercells and the
SC6H4 supercell. The band structures of the H-cluster decorated nanotubes 
with different supercell lengths follow roughly those of the pristine nanotube.
However, there is an important qualitative difference. In the case of
the SC6H4 supercell a band gap opens around the Fermi energy. The
size of the band gap increases with the strength of the perturbation,
e.g., as the number of the hydrogen atoms in the cluster increases. 
Further calculations with supercells containing several H-atom clusters
show that such a band gap opening happens for all supercells in which 
the relative positions of the adjacent adsorbate clusters, or more 
generally bi-site perturbations, fulfill the condition
\begin{equation}\label{condition}
\vec{R} = p\vec{a} + q\vec{b}, \hspace{35 pt} p-q=3M, \hspace{35 pt}
| M\in\mathbb{Z},
\end{equation}
where $\vec{a}$ and $\vec{b}$ are the unit vectors given 
in Fig.~\ref{fig:nanotube}. Thus, the clusters may be in different 
positions on the planes perpendicular to the tube axis as, for example,
the different clusters in Fig.~\ref{fig:nanotube}. Eq. (1)
is actually the condition that a pristine nanotube is metallic 
\cite{Dresselhaus}. Moreover, it was found by Garc\'\i{}a-Lastra 
{\em et al.}~\cite{main} to give also the rule that
two molecules adsorbed on same sublattice sites of a SWNT leave
one of the two transmission channels unaffected.  

The above-mentioned metal to semiconductor transition can be understood 
as follows. When the length of the unit cell of a pristine armchair SWNT 
is artificially tripled the energy band crossing point at the Fermi level 
(corresponding to a $K$ point in the first Brillouin zone of graphene) 
is folded close to the $\Gamma$-point (see Fig.~\ref{fig:bandstructure}(a) 
for Brillouin zone folding lines). Therefore, perturbations with this 
periodicity affect the bandstructure in the same way as the lattice potential 
in the nearly free electron model and open up a band gap around the 
Fermi level. The same happens for any periodicity length of 3M times 
the unit cell length and also when the condition of Eq. (1) is fulfilled. 
But for the periodicity lengths of 3M+1 or 3M+2 times the unit cell
length the band crossing point does not coincide with a reciprocal
lattice boundary and the armchair SWNT remains metallic even with the
bi-site perturbations.

\begin{figure}
\includegraphics[scale=0.35,clip]{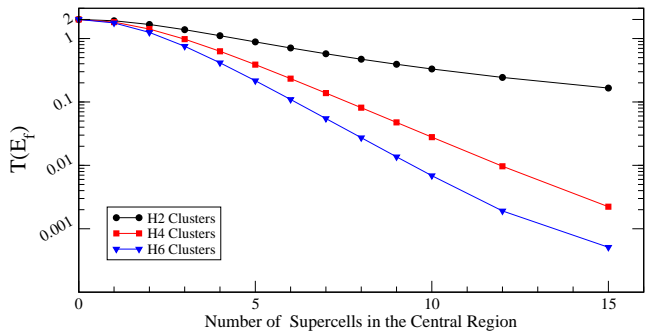}
\caption{\label{fig:FermiTrans} (Color online) Fermi-energy transmission
coefficient for N(SC6H2), N(SC6H4), and N(SC6H6) scattering regions as 
a function of N, the number of the perturbing supercells.
}
\end{figure}

To get another viewpoint about the origin of the band gap opening 
for certain periodic perturbations and to see how periodic 
bi-site perturbations of finite lengths affect the transmission
of the armchair SWCNT, we calculate the transmission functions of several
nanotubes with a varying number of perturbed supercells in the central
region. With increasing number of atoms, the use of DFT for transport 
calculations becomes prohibitively cpu-time consuming and  therefore 
we employ a simple tight-binding method benchmarked against the DFT
results. The upper and lower panels of Fig.~\ref{fig:transport} show 
the transmission functions for the N(SC5H4) and N(SC6H4) central region
systems, respectively. The lengths of the periodic regions increase
from left to right with N=1, 2, 5 and 10. It can be clearly seen
that the transmission in the N(SC6H4) systems decays and sharpens 
in the shape around the Fermi energy with the length of the periodically 
perturbed region.

In contrast, for the N(SC5H4) systems, the transmission around the Fermi energy
remains very close to that of the pristine armchair SWCNT and actually the 
several scatterings develop a nearly constant transmission plateau 
with the increasing length of the periodically perturbed region.

We plot in Fig.~\ref{fig:FermiTrans} the transmission coefficient 
at the Fermi energy for the N(SC6H2), N(SC6H4), and N(SC6H6)  central 
region systems as a function of N. Although the stability of this
H6 cluster has not been established in contrast to H2 and H4 clusters
\cite{clustering}, we use H6 cluster to qualitatively 
mimic the effect of stronger perturbations which cover a full hexagon. 
For long periodically perturbed
regions, a nearly exponential decay can be seen in all three cases
reflecting tunneling through a semiconducting region. In a natural
manner, the decay rate increases strongly as a function of the cluster size. 

\begin{figure}
\includegraphics[scale=0.35,clip]{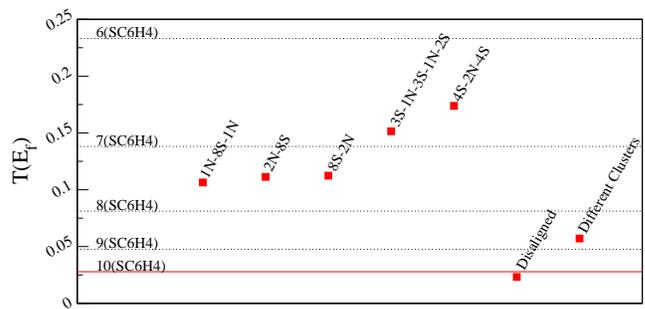}
\caption{\label{fig:Vary} (Color online) Fermi-energy transmission
coefficient for different systems made by manipulating the original 10(SC6H4) 
system by displacing two of the H4 clusters. In the notation giving the 
sequence of the clusters, S and N stand for clusters $Satisfying$ 
and $Not-satisfying$ the condition of Eq. (1). 
Sample results due to changing clusters' positions $(Disaligned)$
around the circumference perpendicular to the tube axis
and due to replacing some of the H4 clusters by H6 or H2 clusters $(Different$ $Clusters)$ 
are also shown. The solid (red) line is the Fermi-energy transmission value 
for the original 10(SC6H4)system and the (black) dotted lines are,
for comparison, those for the N(SC6H4) systems with N=9, 8, 7 and 6.}
\end{figure}

We describe the above phenomena as follows. When the periodic perturbations 
occur with the separation of $n \lambda_F/2$, where $n$ is an integer, all 
the backscattered electron waves at the Fermi level interfere constructively 
suppressing the transmission. As depicted in Fig.~\ref{fig:nanotube} 
the Fermi wavelength of an armchair SWCNT is $3a$ and therefore the 
constructive interference of the backscattering waves takes place for periodic 
central regions constructed, for example, from the SC6H4 supercells but not
for those containing, for example, SC5H4 supercells (See Fig.~\ref{fig:nanotube}).

Next, we explore with the tight-binding model the generality of 
our finding of the opening of the transmission gap at the Fermi level. 
We take the 10(SC6H4) central region system and change the positions of two 
of the H4 clusters so that they don't satisfy Eq. (\ref{condition}). Depending 
on, which two of the clusters one chooses, the transmission changes, but not 
in a radical way. In all combinations the general trend of the decreasing 
transmission is conserved. In Fig.~\ref{fig:Vary} the transmission at the
Fermi level is shown for some of the combinations studied. The transmission is 
close to that of the 8(SC6H4) central region system when all the clusters
satisfying Eq. (\ref{condition}) are adjacent to each others. When the 
clusters satisfying Eq. (\ref{condition}) are separated the transmission is 
close to that of the 7(SC6H4) central region system.

The effects of randomness in the size of the clusters as well as in their positions 
around the circumference perpendicular to the tube axis are also studied in 
the case of the 10(SC6H4) system. First we replace some of the H4 clusters 
with the H2 or H6 clusters and see that the transmission at the Fermi level stays 
low although the exact value depends on the particular combination
in question. Next, in a set of separate calculations, we move the H4 clusters 
around the nanotube on the same perpendicular plane. We find that as long 
as the clusters fulfill Eq. (1), the same destructive effect on 
transmission occurs. Sample results labeled as \textit{Different Clusters} 
and \textit{Disaligned} are included in Fig.~\ref{fig:Vary}. Our findings 
show that the exact periodicity of the clusters is not playing the main role and
the cluster species and their position around the tube may vary, for example,
as depicted in Fig.~\ref{fig:nanotube}. 

Finally, we emphasize that our qualitative findings are valid regardless of 
the type of perturbations beyond the adsorbate clusters. For instance, our 
calculations for carbon nanobuds \cite{main-nanobud, PhysRevB.80.035427}, which 
can be viewed as an example of perturbing a full hexagon, show exactly the same behavior.
Moreover, according to our calculations also different 
hydrogenated vacancy clusters result in the same phenomenon.

\section{Conclusions}

We have studied the effect of multiple bi-site perturbations 
on electronic and transport properties of armchair nanotubes. Our 
calculations show that following a certain relative-position condition, 
a naturally metallic nanotube can turn into semiconducting. The phenomenon 
shows robustness against variations in the types of perturbing species
and also to some extent in their positions. The phenomenon is proposed 
as a means for creating single-SWCNT electronic devices.

\section{Acknowledgment} 
The work has been supported by Finnish Academy through their Center of
Excellence program. We are thankful to R.M. Nieminen, A-P Jauho and
Maria Ganchenkova for useful discussions.

\bibliography{Periodic_Cluster}

\end{document}